\documentclass[10pt]{article}
\usepackage{amsmath}
\usepackage{amssymb}
\usepackage{graphicx}
\usepackage{cite}
\usepackage{color}
\usepackage[labelfont=bf,labelsep=period,justification=raggedright]{caption}

\makeatletter
\renewcommand{\@biblabel}[1]{\quad#1.}
\makeatother
\newcommand{\be}{\begin{equation}}
\newcommand{\ee}{\end{equation}}
\newcommand{\bea}{\begin{eqnarray}}
\newcommand{\eea}{\end{eqnarray}}

\input{tcilatex}
\begin{document}

\date{}

% Title must be 150 characters or less

\begin{flushleft}
{\Large \textbf{Analyzing three-player quantum games in an EPR type setup} } 
% Insert Author names, affiliations and corresponding author email.
\newline
James M.~Chappell$^{1,2,\ast}$, Azhar Iqbal$^{2}$, Derek Abbott$^{2}$ 
\newline
\textbf{1} School of Chemistry and Physics, University of Adelaide, South Australia,
Australia \newline
\textbf{2} School of Electrical and Electronic Engineering, University of
Adelaide, South Australia, Australia \newline
$\ast$ E-mail: james.m.chappell@adelaide.edu.au
\end{flushleft}

% Please keep the abstract between 250 and 300 words

\section*{Abstract}

We use the formalism of Clifford Geometric Algebra (GA) to develop an analysis of
quantum versions of three-player non-cooperative games. The quantum games we
explore are played in an Einstein-Podolsky-Rosen (EPR) type setting. In this
setting, the players' strategy sets remain identical to the ones in the
mixed-strategy version of the classical game that is obtained as a proper
subset of the corresponding quantum game. Using GA we investigate the
outcome of a realization of the game by players sharing GHZ state, W state,
and a mixture of GHZ and W states. As a specific example, we study the game
of three-player Prisoners' Dilemma.

% Please keep the Author Summary between 150 and 200 words
% Use first person. PLoS ONE authors please skip this step. 
% Author Summary not valid for PLoS ONE submissions.   
%\section*{Author Summary}

\section*{Introduction}

The field of game theory \cite{Binmore,Rasmusen} has a long history \cite%
{AbbottDavies}, but was first formalized in 1944 with the work of von
Neumann and Morgenstern \cite{vonNeumanMorgenstern}, aiming to develop
rational analysis of situations that involve strategic interdependence.

Classical game theory has found increasing expression in the field of
physics \cite{AbbottDavies} and its extension to the quantum regime \cite%
{Peres} was proposed by Meyer \cite{MeyerDavid} and Eisert et al \cite{Eisert1999},
though its origins can be traced to earlier works \cite%
{Blaquiere,Wiesner,Mermin,Mermin1}. Early studies in the area of quantum
games focused on the two-player two-strategy non-cooperative games, with the
proposal for a quantum Prisoners' Dilemma (PD) being well known \cite{Eisert1999}.
A natural further development of this work was its extension to multiplayer
quantum games that was explored by Benjamin and Hayden \cite{BenjaminHayden}%
. Du et al. \cite{Du2001,Du2003} explored the phase transitions in quantum
games for the first time that are central in the present article.

The usual approach in three-player quantum games considers players sharing a
three-qubit quantum state with each player accessing their respective qubit
in order to perform local unitary transformation. Quantum games have been
reported \cite{YongJianHan2002} in which players share Greenberger-Horne-Zeilinger
(GHZ) states and the W states \cite{Peres}, while other works have, for
instance, investigated the effects of noise \cite{FlitneyAbbott2,Ramzan} and
the benefits of players forming coalitions \cite{IqbalToor3,FlitneyGreentree}%
.

A suggested approach \cite{IqbalWeigert,IqbalEpr:2005,IqbalCheon,IqbalCheonAbbott} in constructing
quantum games uses an Einstein-Podolsky-Rosen (EPR) type setting \cite%
{EPR,Bohm,Bell,Bell1,Bell2,Aspect,ClauserShimony,Cereceda}. In this
approach, quantum games are setup with an EPR type apparatus, with the
players' strategies being local actions related to their qubit, consisting
of a linear combination (with real coefficients) of (spin or polarization)
measurements performed in two selected directions.

Note that in a standard arrangement for playing a mixed-strategy game,
players are faced with the identical situation, in that in each run, a
player has to choose one out of two pure strategies. As the players'
strategy sets remain classical, the EPR type setting avoids a well known
criticism \cite{EnkPike2002} of quantum games. This criticism refers to
quantization procedures in which players are given access to extended
strategy sets, relative to what they are allowed to have in the classical
game. Quantum games constructed with an EPR type setting have been studied
in situations involving two players \cite{IqbalCheon} and also three players 
\cite{IqbalCheonAbbott}. The applications of three-player quantum games
include describing three-party situations, involving strategic interaction
in quantum communication \cite{NielsenChuang:2002}.

In recent works, the formalism of Clifford's geometric algebra (GA) \cite%
{GA,GA1,Doran:2003,Venzo2007,Dorst:2002} has been applied to
the analysis of two-player quantum games with significant benefits \cite%
{CIL,ChappellB}, and so is also adopted here in the analysis of
three-player quantum games. The use of GA is justified on the grounds that
the Pauli spin algebra is a matrix representation of Clifford's geometric
algebra in $\mathcal{R}^{3}$, and hence we are choosing to work directly
with the underlying Clifford algebra. There are also several other
documented benefits of GA such as:

a) The unification of the dot and cross product into a single product, has
the significant advantage of possessing an inverse. This results in
increased mathematical compactness, thereby aiding physical intuition and
insight \cite{Boudet}.

b) The use of the Pauli and Dirac matrices also unnecessarily introduces the
imaginary scalars, in contrast to GA, which uses exclusively real elements 
\cite{Hestenes}. This fact was also pointed out by Sommerfield in 1931, who
commented that \textit{\ `Dirac's use of matrices simply rediscovered
Clifford algebra'} \cite{Schmeikal}.

c) In the density matrix formalism of quantum mechanics, the expectation for
an operator $Q$ is given by $\text{Tr}(\rho Q)=\langle \psi |Q|\psi \rangle $%
, from which we find the isomorphism to GA, \textrm{Tr}$(\rho
Q)\leftrightarrow \langle \rho Q\rangle _{0}$, the subscript zero,
indicating to take the scalar part of the algebraic product $\rho Q$, where $%
\rho $ and $Q$ are now constructed from real Clifford elements. This leads
to a uniquely compact expression for the overlap probability between two
states in the $N$-particle case, given by Eq.~(\ref{eq:DoranOverlapProb}),
which allows straightforward calculations that normally require $8\times 8$
complex matrices representing operations on three qubits.

d) Pauli wave functions are isomorphic to the quaternions, and hence
represent rotations of particle states \cite{Horn}. This fact paves the way
to describe general unitary transformations on qubits, in a simplified
algebraic form, as \textit{rotors}. In regard to Hestenes' analysis of the
Dirac equation using GA, Boudet \cite{Boudet} notes that, `the use of the
pure real formalism of Hestenes brings noticeable simplifications and above
all the entire geometrical clarification of the theory of the electron.'

e) Recent works \cite{CIL,ChappellB,MeyerDavid} show that GA provides a
better intuitive understanding of Meyer's quantum penny flip game \cite%
{MeyerDavid}, using operations in $3$-space with \textit{real coordinates},
permitting helpful visualizations in determining the quantum player's
winning strategy. Also, Christian \cite{Christian,christian2011restoring} has recently used GA to
produce thought provoking investigations into some of the foundational
questions in quantum mechanics.

Our quantum games use an EPR type setting and players have access to general
pure quantum states. We determine constraints that ensure a faithful
embedding of the mixed-strategy version of the original classical game
within the corresponding quantum game. We find how a Pareto-optimal quantum
outcome emerges in three-player quantum PD game at high entanglement. We
also report phase transitions taking place with increasing entanglement when
players share a mixture of GHZ and W type states in superposition.

In an earlier paper \cite{IqbalCheonAbbott}, two of the three authors contributed to developing
an entirely probabilistic framework for the analysis of three-player quantum
games that are also played using an EPR type setting, whereas the present paper, though
using an EPR type setting, provides an analysis from the perspective of quantum mechanics, 
with the mathematical formalism of GA. The previous work analyzed quantum games from the non-factorizable property of a joint
probability distribution relevant to a physical system that the players
shared in order to implement the game. For the game of three-player
Prisoners' Dilemma, our probabilistic analysis showed that
non-factorizability of a joint probability distribution indeed can lead to a
new equilibrium in the game. The three-player quantum Prisoners' Dilemma,
in the present analysis, however, moves to the next step and explores the phase
structure relating players' payoffs with shared entanglement and also the
impact of players sharing GHZ and W states and their mixture. We believe
that without using the powerful formalism of GA, a similar analysis will
nearly be impossible to perform using an entirely probabilistic approach as
developed in \cite{IqbalCheon}.

\section*{EPR setting for playing quantum games}

The EPR setting \cite{IqbalWeigert,IqbalCheon,IqbalCheonAbbott} two player
quantum games involves a large number of runs when, in a run, two halves of
an EPR pair originate from the same source and move in the opposite
directions. Player Alice receives one half whereas player Bob receives the
other half. To keep the non-cooperative feature of the game, it is assumed
that players Alice and Bob are located at some distance from each other and
are not unable to communicate between themselves. The players, however, can
communicate about their actions, which they perform on their received
halves, to a \textit{referee} who organizes the game and ensures that the
rules of the game are followed. The referee makes available two directions
to each player. In a run, each player has to choose one of two available
directions. The referee rotates Stern-Gerlach type detectors \cite{Peres}
along the two chosen directions and performs quantum measurement. The
outcome of the quantum measurement, on Alice's side, and on Bob's side of
the Stern-Gerlach detectors, is either $+1$ or $-1$. Runs are repeated as
the players receive a large number of halves in pairs, when each pair comes
from the same source and the measurement outcomes are recorded for all runs.
A player's strategy, defined over a large number of runs, is a linear
combination (with normalized and real coefficients) of the two directions
along which the measurement is performed. The referee makes public the
payoff relations at the start of the game and announces rewards to the
players after the completion of runs. The payoff relations are constructed
in view of a) the matrix of the game, b) the list of players' choices of
directions over a large number of runs, and c) the list of measurement
outcomes that the referee prepares using his/her Stern-Gerlach apparatus.

For a three-player quantum game, this setting is extended to consider three
players Alice, Bob and Chris who are located at the three arms of an EPR
system \cite{Peres}. In the following they will be denoted by $A,$ $B$ and $C
$, respectively. As it is the case with two-player EPR setting, in a run of
the experiment, each player chooses one out of two directions.

We have used the EPR setting in view of the well known Enk and Pike's
criticism \cite{EnkPike2002} of quantum games that are played using Eisert et
al's setting \cite{Eisert1999}. Essentially this criticism attempts to equate a
quantum game to a classical game in which the players are given access to an
extended set of classical strategies. The present paper uses an EPR setting
in which each player has two classical strategies consisting of the two
choices he/she can make between two directions along which a quantum
measurement can be performed. That is, the player's pure strategy, in a run,
consists of choosing one direction out of the two. As the sets of strategies
remain exactly identical in both the classical and the quantum forms of the
game, it is difficult to construct an Enk and Pike type argument for a
quantum game that is played with an EPR setting.

\begin{figure}[!ht]
\begin{center}
\includegraphics[width=3.9in]{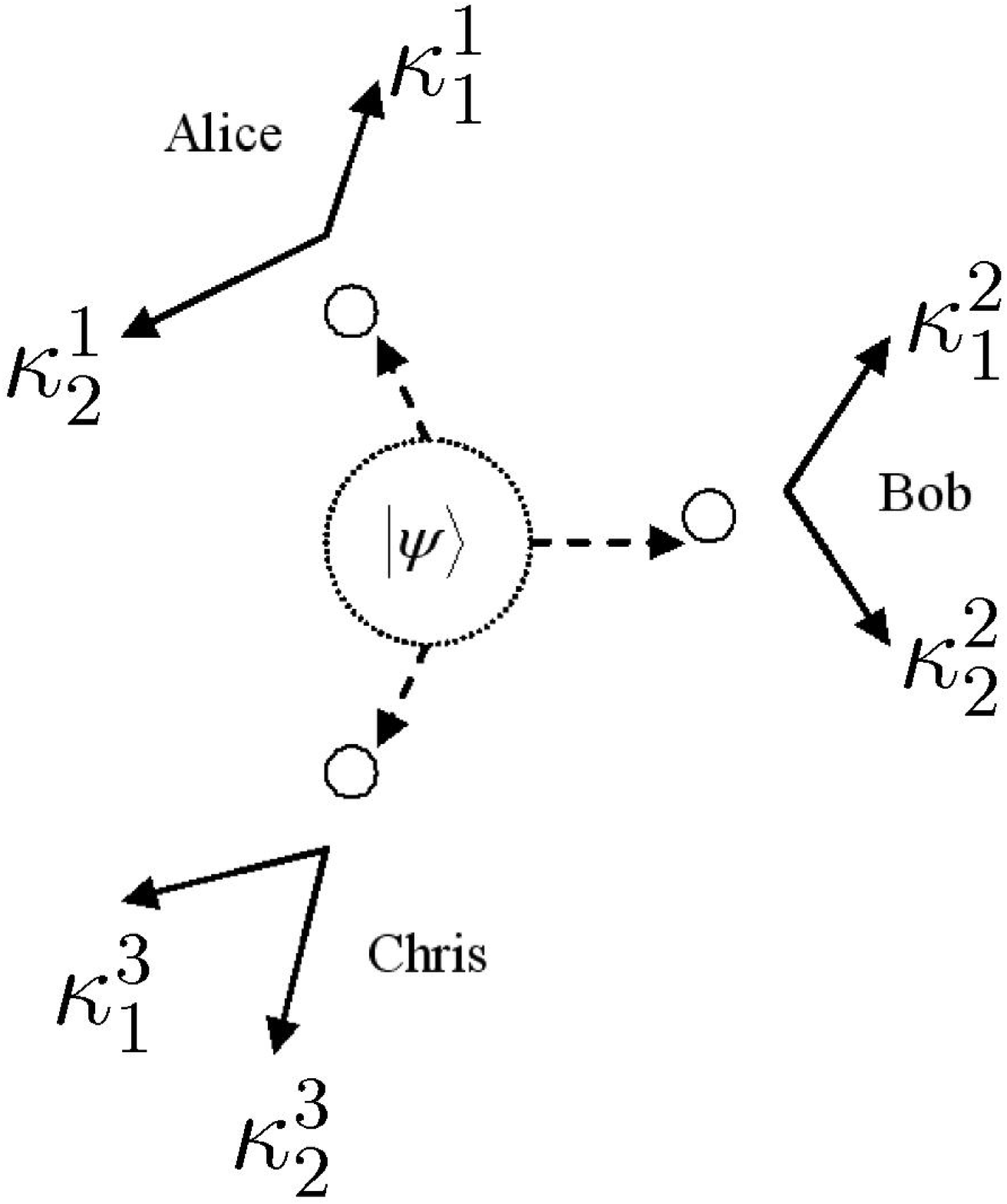}
\end{center}
\caption{{\bf{The EPR setup for three-player quantum game.}} A three-qubit
entangled quantum state is distributed to the three players, who each choose
between two possible measurement directions.}
\label{eprFigure}
\end{figure}

As Fig.~\ref{eprFigure} shows, we represent Alice's two directions as $%
\kappa _{1}^{1},\kappa _{2}^{1}$. Similarly, Bob's directions are $\kappa
_{1}^{2},\kappa _{2}^{2}$ and Chris' are $\kappa _{1}^{3},\kappa _{2}^{3}$.
The players measurement directions form a triplet out of eight possible
cases $(\kappa _{1}^{1},\kappa _{1}^{2},\kappa _{1}^{3}),$ $(\kappa
_{1}^{1},\kappa _{2}^{2},\kappa _{1}^{3}),$ $(\kappa _{2}^{1},\kappa
_{1}^{2},\kappa _{1}^{3}),$ $(\kappa _{2}^{1},\kappa _{2}^{2},\kappa
_{1}^{3}),$ $(\kappa _{1}^{1},\kappa _{1}^{2},\kappa _{2}^{3}),$ $(\kappa
_{1}^{1},\kappa _{2}^{2},\kappa _{2}^{3}),$ $(\kappa _{2}^{1},\kappa
_{1}^{2},\kappa _{2}^{3}),$ $(\kappa _{2}^{1},\kappa _{2}^{2},\kappa
_{2}^{3})$ and measurement is performed along the chosen directional
triplet. The measurement outcome for each player along their chosen
direction is $+1$ or $-1$.

% Location for Figure 1

Over a large number of runs the players sequentially receive three-particle
systems emitted from a source and a record is maintained of the players'
choices of directions over all runs. One of the eight possible outcomes $%
(+1,+1,+1),$ $(+1,-1,+1),$ $(-1,+1,+1),$ $(-1,-1,+1),$ $(+1,+1,-1),$ $%
(+1,-1,-1),$ $(-1,+1,-1),$ $(-1,-1,-1)$ emerges out of the measurement in an
individual run, with the first entry for Alice's outcome, the second entry
for Bob's outcome and the third entry for Chris' outcome.

In the following we express the players' payoff relations in terms of the
outcomes of these measurements. These payoffs depend on the triplets of the
players' strategic choices made over a large number of runs and on the
dichotomic outcomes of the measurements performed along those directions.

\subsection*{Players' sharing a symmetric initial state}

We consider the situation in which an initial quantum state of three qubits
is shared among three players. To obtain a fair game, we assume this state
is symmetric with regard to the interchange of the three players. The GHZ
state is a natural candidate given by 
\begin{equation}
|\text{GHZ}\rangle =\cos \frac{\gamma }{2}|000\rangle +\sin \frac{\gamma }{2}%
|111\rangle ,
\end{equation}%
where we have an entanglement angle $\gamma \in \Re $, which has been shown 
\cite{Peres} to be capable of producing the maximally entangled three qubit
state. Alternatively we could start with the W entangled state 
\begin{equation}
|\text{W}\rangle =\frac{1}{\sqrt{3}}(|100\rangle +|010\rangle +|001\rangle ).
\end{equation}%
The other symmetric state would be an inverted W state 
\begin{equation}
|\bar{\text{W}}\rangle =\frac{1}{\sqrt{3}}(|110\rangle +|011\rangle
+|101\rangle ).  \label{W2State}
\end{equation}

After the measurement along three directions selected by the players, each
player is rewarded according to a payoff matrix $G^{\mathcal{P}}$, for each
player $\mathcal{P}\in \{A,B,C\}$. Thus the expected payoffs for a player is
given by 
\begin{equation}
\Pi _{\mathcal{P}}(\kappa ^{1},\kappa ^{2},\kappa
^{3})=\sum_{i,j,k=0}^{1}G_{ijk}^{\mathcal{P}}P_{ijk},  \label{eq:AlicePayoff}
\end{equation}%
where $P_{ijk}$ is the probability the state $|i\rangle |j\rangle |k\rangle $
is obtained after measurement, with $i,j,k\in \{0,1\}$, along the three
directions $\kappa ^{1},\kappa ^{2},\kappa ^{3}$ chosen by Alice, Bob and
Chris respectively. In the EPR setting, $\kappa ^{1}$ can be either of
Alice's two directions i.e. $\kappa _{1}^{1}$ or $\kappa _{2}^{1}$ and
similarly for Bob and Chris.

\subsection*{Clifford's geometric algebra}

The formalism of GA \cite%
{GA,GA1,Doran:2003,Venzo2007,Dorst:2002} has been shown to
provide an equivalent description to the conventional tensor product
formalism of quantum mechanics.

To set up the GA framework for representing quantum states, we begin by
defining $\sigma _{1},\sigma _{2},\sigma _{3}$ as a right-handed set of
orthonormal basis vectors, with 
\begin{equation}
\sigma _{i}.\sigma _{j}=\delta _{ij},
\end{equation}%
where $\delta _{ij}$ is Kronecker delta. Multiplication between algebraic
elements is defined to be the geometric product, which for two vectors $u$
and $v$ is given by

\begin{equation}
uv=u.v+u\wedge v,
\end{equation}%
where $u.v$ is the conventional symmetric dot product and $u\wedge v$ is the
anti-symmetric outer product related to the Gibb's cross product by $u\times
v=-\iota u\wedge v$, where $\iota =\sigma _{1}\sigma _{2}\sigma _{3}$. For
distinct basis vectors we find 
\begin{equation}
\sigma _{i}\sigma _{j}=\sigma _{i}.\sigma _{j}+\sigma _{i}\wedge \sigma
_{j}=\sigma _{i}\wedge \sigma _{j}=-\sigma _{j}\wedge \sigma _{i}=-\sigma
_{j}\sigma _{i}.
\end{equation}%
This can be summarized by 
\begin{equation}
\sigma _{i}\sigma _{j}=\delta _{ij}+\iota \epsilon _{ijk}\sigma _{k},
\end{equation}%
where $\epsilon _{ijk}$ is the Levi-Civita symbol. We can therefore see that 
$\iota $ squares to minus one, that is $\iota ^{2}=\sigma _{1}\sigma
_{2}\sigma _{3}\sigma _{1}\sigma _{2}\sigma _{3}=\sigma _{1}\sigma
_{2}\sigma _{1}\sigma _{2}=-1$ and commutes with all other elements and so
has identical properties to the unit imaginary $i$.
Thus we have an isomorphism between the basis vectors $\sigma _{1},\sigma
_{2},\sigma _{3}$ and the Pauli matrices through the use of the geometric
product.

In order to express quantum states in GA we use the one-to-one mapping \cite%
{Doran:2003,Dorst:2002} defined as follows 
\begin{equation}
|\psi \rangle =\alpha |0\rangle +\beta |1\rangle =%
\begin{bmatrix}
a_{0}+\mathrm{i}a_{3} \\ 
-a_{2}+\mathrm{i}a_{1}%
\end{bmatrix}%
\leftrightarrow \psi =a_{0}+a_{1}\iota \sigma _{1}+a_{2}\iota \sigma
_{2}+a_{3}\iota \sigma _{3},  \label{eq:spinorMapping}
\end{equation}%
where $a_{i}$ are real scalars.

For a single particle we then have the basis vectors 
\begin{equation}
|0\rangle \longleftrightarrow 1,\text{ \ \ }|1\rangle \longleftrightarrow -{%
\iota }\sigma _{2}
\end{equation}%
and so for three particles we can use as a basis 
\begin{subequations}
\label{threeQubitBasisGA}
\begin{eqnarray}
|0\rangle |0\rangle |0\rangle &\longleftrightarrow &1 \\
|0\rangle |0\rangle |1\rangle &\longleftrightarrow &-{\iota }\sigma _{2}^{3}
\\
|0\rangle |1\rangle |0\rangle &\longleftrightarrow &-{\iota }\sigma _{2}^{2}
\\
|0\rangle |1\rangle |1\rangle &\longleftrightarrow &{\iota }\sigma _{2}^{2}{%
\iota }\sigma _{2}^{3}, \\
|1\rangle |0\rangle |0\rangle &\longleftrightarrow &-{\iota }\sigma _{2}^{1}
\\
|1\rangle |0\rangle |1\rangle &\longleftrightarrow &{\iota }\sigma _{2}^{1}{%
\iota }\sigma _{2}^{3} \\
|1\rangle |1\rangle |0\rangle &\longleftrightarrow &{\iota }\sigma _{2}^{1}{%
\iota }\sigma _{2}^{2} \\
|1\rangle |1\rangle |1\rangle &\longleftrightarrow &-{\iota }\sigma _{2}^{1}{%
\iota }\sigma _{2}^{2}{\iota }\sigma _{2}^{3},
\end{eqnarray}%
where to reduce the number of superscripts representing particle number we
write ${\iota }^{1}\sigma _{2}^{1}$ as ${\iota }\sigma _{2}^{1}$. General
unitary operations are equivalent to rotors in GA \cite{Doran:2003},
represented as 
\end{subequations}
\begin{equation}
R(\theta _{1},\theta _{2},\theta _{3})=\mathrm{e}^{-\theta _{3}{\iota }%
\sigma _{3}/2}\mathrm{e}^{-\theta _{1}{\iota }\sigma _{2}/2}\mathrm{e}%
^{-\theta _{2}{\iota }\sigma _{3}/2},  \label{eq:GenUnitaryRotation}
\end{equation}%
which is in Euler angle form and can completely explore the available space
of a single qubit. Using the definition of unitary operations given by Eq.~(%
\ref{eq:GenUnitaryRotation}) we define $A=R(\alpha _{1},\alpha _{2},\alpha
_{3}),$ $B=R(\beta _{1},\beta _{2},\beta _{3}),$ $C=R(\chi _{1},\chi
_{2},\chi _{3})$ for general unitary transformations acting locally on each
of the three players qubit in order to generalize the starting state, that
is the GHZ or W states, as far as possible.

We define a separable state $\phi =KLM$, where $K,$ $L$ and $M$ are single
particle rotors, which allow the players' measurement directions to be
specified on the first, second and third qubit respectively. The state to be
measured is now projected onto this separable state $\phi $. The overlap
probability between two states $\psi $ and $\phi $ in the $N$-particle case
is given in Ref. \cite{Doran:2003} as 
\begin{equation}
P(\psi ,\phi )=2^{N-2}[\langle \psi E\psi ^{\dagger }\phi E\phi ^{\dagger
}\rangle _{0}-\langle \psi J\psi ^{\dagger }\phi J\phi ^{\dagger }\rangle
_{0}],  \label{eq:DoranOverlapProb}
\end{equation}%
where the angle brackets $\langle \rangle _{0}$ mean to retain only the
scalar part of the expression and $E$ and $J$ are defined for 3 particles in
Ref. \cite{Doran:2003} as 
\begin{subequations}
\begin{eqnarray}
E &=&\prod_{i=2}^{N}\frac{1}{2}(1-{\iota }\sigma _{3}^{1}{\iota }\sigma
_{3}^{i})=\frac{1}{4}(1-{\iota }\sigma _{3}^{1}{\iota }\sigma _{3}^{2}-{%
\iota }\sigma _{3}^{1}{\iota }\sigma _{3}^{3}-{\iota }\sigma _{3}^{2}{\iota }%
\sigma _{3}^{3}) \\
J &=&E{\iota }\sigma _{3}^{1}=\frac{1}{4}({\iota }\sigma _{3}^{1}+{\iota }%
\sigma _{3}^{2}+{\iota }\sigma _{3}^{3}-{\iota }\sigma _{3}^{1}{\iota }%
\sigma _{3}^{2}{\iota }\sigma _{3}^{3}).
\end{eqnarray}

The $\dagger $ operator acts the same as complex conjugation: flipping the
sign of $\iota $ and inverting the order of the terms.

% Results and Discussion can be combined.

\section*{Results}

We now, firstly, calculate the observables from Eq.~(\ref{eq:DoranOverlapProb}) for
the GHZ state in GA, which from Eq.~(\ref{threeQubitBasisGA}) gives 
\end{subequations}
\begin{equation}
\psi =ABC(\cos \frac{\gamma }{2}-\sin \frac{\gamma }{2}{\iota }\sigma
_{2}^{1}{\iota }\sigma _{2}^{2}{\iota }\sigma _{2}^{3}),  \label{GHZ in GA}
\end{equation}%
where $A,$ $B,$ and $C$ represent the referee's local unitary actions,
written as rotors $A,$ $B,$ and $C$ in GA, on the respective player's
qubits, in order to generalize the starting state. Referring to Eq.~(\ref%
{eq:DoranOverlapProb}), we firstly calculate 
\begin{align}
\psi J\psi ^{\dagger }& =\frac{1}{4}ABC(\cos \frac{\gamma }{2}-\sin \frac{%
\gamma }{2}{\iota }\sigma _{2}^{1}{\iota }\sigma _{2}^{2}{\iota }\sigma
_{2}^{3})({\iota }\sigma _{3}^{1}+{\iota }\sigma _{3}^{2}+{\iota }\sigma
_{3}^{3}-{\iota }\sigma _{3}^{1}{\iota }\sigma _{3}^{2}{\iota }\sigma
_{3}^{3})  \notag \\
& \times (\cos \frac{\gamma }{2}+\sin \frac{\gamma }{2}{\iota }\sigma
_{2}^{1}{\iota }\sigma _{2}^{2}{\iota }\sigma _{2}^{3})C^{\dagger
}B^{\dagger }A^{\dagger }  \notag \\
& =\frac{1}{4}ABC(\cos \gamma -\sin \gamma {\iota }\sigma _{2}^{1}{\iota }%
\sigma _{2}^{2}{\iota }\sigma _{2}^{3})({\iota }\sigma _{3}^{1}+{\iota }%
\sigma _{3}^{2}+{\iota }\sigma _{3}^{3}-{\iota }\sigma _{3}^{1}{\iota }%
\sigma _{3}^{2}{\iota }\sigma _{3}^{3})C^{\dagger }B^{\dagger }A^{\dagger } 
\notag \\
& =\frac{1}{4}\cos \gamma (R_{3}+S_{3}+T_{3}-R_{3}S_{3}T_{3})+\sin \gamma
(R_{1}S_{2}T_{2}+R_{2}S_{1}T_{2}+R_{2}S_{2}T_{1}-R_{1}S_{1}T_{1})
\end{align}%
where $R_{k}={\iota }A\sigma _{k}A^{\dagger },\,S_{k}={\iota }B\sigma
_{k}B^{\dagger },\,T_{k}={\iota }C\sigma _{k}C^{\dagger }$. We also
calculate 
\begin{eqnarray}
\psi E\psi ^{\dagger } &=&\frac{1}{4}ABC(\cos \frac{\gamma }{2}-\sin \frac{%
\gamma }{2}{\iota }\sigma _{2}^{1}{\iota }\sigma _{2}^{2}{\iota }\sigma
_{2}^{3})(1-{\iota }\sigma _{3}^{1}{\iota }\sigma _{3}^{2}-{\iota }\sigma
_{3}^{1}{\iota }\sigma _{3}^{3}-{\iota }\sigma _{3}^{2}{\iota }\sigma
_{3}^{3})  \notag \\
&\times &(\cos \frac{\gamma }{2}+\sin \frac{\gamma }{2}{\iota }\sigma
_{2}^{1}{\iota }\sigma _{2}^{2}{\iota }\sigma _{2}^{3})C^{\dagger
}B^{\dagger }A^{\dagger }  \notag \\
&=&\frac{1}{4}ABC\left( 1-{\iota }\sigma _{3}^{1}{\iota }\sigma _{3}^{2}-{%
\iota }\sigma _{3}^{1}{\iota }\sigma _{3}^{3}-{\iota }\sigma _{3}^{2}{\iota }%
\sigma _{3}^{3}\right) C^{\dagger }B^{\dagger }A^{\dagger }  \notag \\
&=&\frac{1}{4}\left( 1-R_{3}S_{3}-R_{3}T_{3}-S_{3}T_{3}\right) .
\end{eqnarray}%
For measurement defined with $K=e^{-{\iota }\kappa \sigma _{2}^{1}/2}$, $%
L=e^{-{\iota }\kappa \sigma _{2}^{2}/2}$ and $M=e^{-{\iota }\kappa \sigma
_{2}^{3}/2}$ allowing a rotation of the detectors by an angle $\kappa $,
where we have written $\kappa ^{1}\sigma _{2}^{1}$ as $\kappa \sigma
_{2}^{1} $, we find 
\begin{subequations}
\begin{eqnarray}
\phi J\phi ^{\dagger } &=&\frac{1}{4}({\iota }\sigma _{3}^{1}e^{{\iota }%
\kappa \sigma _{2}^{1}}+{\iota }\sigma _{3}^{2}e^{{\iota }\kappa \sigma
_{2}^{2}}+{\iota }\sigma _{3}^{3}e^{{\iota }\kappa \sigma _{2}^{3}}-{\iota }%
\sigma _{3}^{1}{\iota }\sigma _{3}^{2}{\iota }\sigma _{3}^{3}e^{{\iota }%
\kappa \sigma _{2}^{1}}e^{{\iota }\kappa \sigma _{2}^{2}}e^{{\iota }\kappa
\sigma _{2}^{3}}) \\
\phi E\phi ^{\dagger } &=&\frac{1}{4}(1-{\iota }\sigma _{3}^{1}{\iota }%
\sigma _{3}^{2}e^{{\iota }\kappa \sigma _{2}^{1}}e^{{\iota }\kappa \sigma
_{2}^{2}}-{\iota }\sigma _{3}^{1}{\iota }\sigma _{3}^{3}e^{{\iota }\kappa
\sigma _{2}^{1}}e^{{\iota }\kappa \sigma _{2}^{3}}-{\iota }\sigma _{3}^{2}{%
\iota }\sigma _{3}^{3}e^{{\iota }\kappa \sigma _{2}^{2}}e^{{\iota }\kappa
\sigma _{2}^{3}}).
\end{eqnarray}%
From Eq.~(\ref{eq:DoranOverlapProb}) we find 
\end{subequations}
\begin{eqnarray}
2\langle \psi E\psi ^{\dagger }\phi E\phi ^{\dagger }\rangle &=&\frac{1}{8}%
\left( 1-R_{3}S_{3}-R_{3}T_{3}-S_{3}T_{3}\right)  \notag \\
&&\times (1-{\iota }\sigma _{3}^{1}{\iota }\sigma _{3}^{2}e^{{\iota }\kappa
\sigma _{2}^{1}}e^{{\iota }\kappa \sigma _{2}^{2}}-{\iota }\sigma _{3}^{1}{%
\iota }\sigma _{3}^{3}e^{{\iota }\kappa \sigma _{2}^{1}}e^{{\iota }\kappa
\sigma _{2}^{3}}-{\iota }\sigma _{3}^{2}{\iota }\sigma _{3}^{3}e^{{\iota }%
\kappa \sigma _{2}^{2}}e^{{\iota }\kappa \sigma _{2}^{3}})  \notag \\
&=&\frac{1}{8}[1+(-)^{l+m}X(\kappa ^{1})Y(\kappa ^{2})+(-)^{l+n}X(\kappa
^{1})Z(\kappa ^{3})+(-)^{m+n}Y(\kappa ^{2})Z(\kappa ^{3})]  \notag \\
&=&\frac{1}{8}%
[1+(-)^{l+m}X_{i}Y_{j}+(-)^{l+n}X_{i}Z_{k}+(-)^{m+n}Y_{j}Z_{k}],
\label{eq:Method1MeasureE}
\end{eqnarray}%
where $l,m,n\in \{0,1\}$ refers to measuring a $|0\rangle $ or $|1\rangle $
state, respectively, and using the standard results listed in the Appendix,
we have 
\begin{subequations}
\label{eq:XYZRelations}
\begin{eqnarray}
X_{i}=X(\kappa _{i}^{1}) &=&\cos \alpha _{1}\cos \kappa _{i}^{1}+\cos \alpha
_{3}\sin \alpha _{1}\sin \kappa _{i}^{1}, \\
Y_{j}=Y(\kappa _{j}^{2}) &=&\cos \beta _{1}\cos \kappa _{j}^{2}+\cos \beta
_{3}\sin \beta _{1}\sin \kappa _{j}^{2}, \\
Z_{k}=Z(\kappa _{k}^{3}) &=&\cos \chi _{1}\cos \kappa _{k}^{3}+\cos \chi
_{3}\sin \chi _{1}\sin \kappa _{k}^{3},
\end{eqnarray}%
with $i,j,k\in \{1,2\}$, representing the two measurement directions
available to each player. Also from Eq.~(\ref{eq:DoranOverlapProb}) we have 
\end{subequations}
\begin{align}
-2\langle \psi J\psi ^{\dagger }\phi J\phi ^{\dagger }\rangle & =-\frac{1}{8}%
\langle (\cos \gamma (R_{3}+S_{3}+T_{3}-R_{3}S_{3}T_{3})  \notag \\
& +\sin \gamma
(R_{1}S_{2}T_{2}+R_{2}S_{1}T_{2}+R_{2}S_{2}T_{1}-R_{1}S_{1}T_{1}))  \notag \\
& \times ({\iota }\sigma _{3}^{1}e^{{\iota }\kappa \sigma _{2}^{1}}+{\iota }%
\sigma _{3}^{2}e^{{\iota }\kappa \sigma _{2}^{2}}+{\iota }\sigma _{3}^{3}e^{{%
\iota }\kappa \sigma _{2}^{3}}-{\iota }\sigma _{3}^{1}{\iota }\sigma _{3}^{2}%
{\iota }\sigma _{3}^{3}e^{{\iota }\kappa \sigma _{2}^{1}}e^{{\iota }\kappa
\sigma _{2}^{2}}e^{{\iota }\kappa \sigma _{2}^{3}})\rangle _{0}  \notag \\
& =\frac{1}{8}(\cos \gamma
((-)^{l}X_{i}+(-)^{m}Y_{j}+(-)^{n}Z_{k}+(-)^{lmn}X_{i}Y_{j}Z_{k})  \notag \\
& +(-)^{lmn}\sin \gamma
(F_{i}V_{j}W_{k}+U_{i}G_{j}W_{k}+U_{i}V_{j}H_{k}-F_{i}G_{j}H_{k}))  \notag \\
& =\frac{1}{8}[\cos \gamma
\{(-)^{l}X_{i}+(-)^{m}Y_{j}+(-)^{n}Z_{k}+(-)^{lmn}X_{i}Y_{j}Z_{k}%
\}+(-)^{lmn}\sin \gamma \Theta _{ijk}],  \notag \\
&  \label{eq:Method1MeasureJ}
\end{align}%
where 
\begin{subequations}
\label{eq:FGHRelations}
\begin{align}
F_{i}=F(\kappa ^{1})& =-\sin \kappa _{i}^{1}(\cos \alpha _{1}\cos \alpha
_{2}\cos \alpha _{3}-\sin \alpha _{2}\sin \alpha _{3})+\sin \alpha _{1}\cos
\alpha _{2}\cos \kappa _{i}^{1}, \\
G_{j}=G(\kappa ^{2})& =-\sin \kappa _{j}^{2}(\cos \beta _{1}\cos \beta
_{2}\cos \beta _{3}-\sin \beta _{2}\sin \beta _{3})+\sin \beta _{1}\cos
\beta _{2}\cos \kappa _{j}^{2}, \\
H_{k}=H(\kappa ^{3})& =-\sin \kappa _{k}^{3}(\cos \chi _{1}\cos \chi
_{2}\cos \chi _{3}-\sin \chi _{2}\sin \chi _{3})+\sin \chi _{1}\cos \chi
_{2}\cos \kappa _{k}^{3}
\end{align}%
and 
\end{subequations}
\begin{subequations}
\label{eq:UVWRelations}
\begin{align}
U_{i}=U(\kappa ^{1})& =\sin \kappa _{i}^{1}(\cos \alpha _{2}\sin \alpha
_{3}+\sin \alpha _{2}\cos \alpha _{3}\cos \alpha _{1})-\sin \alpha _{1}\sin
\alpha _{2}\cos \kappa _{i}^{1}, \\
V_{j}=V(\kappa ^{2})& =\sin \kappa _{j}^{2}(\cos \beta _{2}\sin \beta
_{3}+\sin \beta _{2}\cos \beta _{3}\cos \beta _{1})-\sin \beta _{1}\sin
\beta _{2}\cos \kappa _{j}^{2}, \\
W_{k}=W(\kappa ^{3})& =\sin \kappa _{k}^{3}(\cos \chi _{2}\sin \chi
_{3}+\sin \chi _{2}\cos \chi _{3}\cos \chi _{1})-\sin \chi _{1}\sin \chi
_{2}\cos \kappa _{k}^{3}
\end{align}%
and 
\end{subequations}
\begin{equation}
\Theta
_{ijk}=F_{i}V_{j}W_{k}+U_{i}G_{j}W_{k}+U_{i}V_{j}H_{k}-F_{i}G_{j}H_{k}.
\label{eq:thetaEquation}
\end{equation}%
So we find from Eq.~(\ref{eq:DoranOverlapProb}) the probability to observe a
particular state after measurement as 
\begin{eqnarray}
P_{lmn} &=&\frac{1}{8}[1+\cos \gamma
\{(-)^{l}X_{i}+(-)^{m}Y_{j}+(-)^{n}Z_{k}\}  \notag \\
&+&(-)^{lm}X_{i}Y_{j}+(-)^{ln}X_{i}Z_{k}+(-)^{mn}Y_{j}Z_{k}+(-)^{lmn}\{\cos
\gamma X_{i}Y_{j}Z_{k}+\sin \gamma \Theta _{ijk}\}].  \notag \\
&&  \label{eq:ThreePlayerFinalDensity}
\end{eqnarray}

For instance, at $\gamma =0$ we obtain 
\begin{equation}
P_{lmn}=\frac{1}{8}(1+(-)^{l}X_{i})(1+(-)^{m}Y_{j})(1+(-)^{n}Z_{k}),
\end{equation}%
which shows a product state, as expected. Alternatively with general
entanglement, but no operation on the third qubit, that is $\chi _{i}=0$, we
have 
\begin{eqnarray}
P_{lm} &=&\frac{1}{8}[1+\cos \gamma
\{(-)^{l}X_{i}+(-)^{m}Y_{j}+1+(-)^{lmn}X_{i}Y_{j}\}  \notag \\
&+&(-)^{lm}X_{i}Y_{j}+(-)^{l}X_{i}+(-)^{m}Y_{j}].  \notag \\
&=&\frac{1}{8}[(1+\cos \gamma )(1+(-)^{l}X_{i})(1+(-)^{m}Y_{j})],
\label{eq:twoQubits}
\end{eqnarray}%
which shows that for the GHZ type entanglement each pair of qubits is
mutually unentangled.

\subsection*{Obtaining the payoff relations}

We extend the approach of Ichikawa and Tsutsui \cite{IchikawaTsutsui} to
three qubits and represent the permutation of signs introduced by the
measurement process. For Alice we define 
\begin{subequations}
\label{eq:aDefs}
\begin{eqnarray}
a_{000} &=&\frac{1}{8}\sum_{ijk}G_{ijk}^{A},\,\,\,\,a_{100}=\frac{1}{8}%
\sum_{ijk}(-)^{i}G_{ijk}^{A},  \label{parametersA} \\
a_{010} &=&\frac{1}{8}\sum_{ijk}(-)^{j}G_{ijk}^{A},\,\,\,\,a_{001}=\frac{1}{8%
}\sum_{ijk}(-)^{k}G_{ijk}^{A},  \label{parametersB} \\
a_{110} &=&\frac{1}{8}\sum_{ijk}(-)^{i+j}G_{ijk}^{A},\,\,\,\,a_{011}=\frac{1%
}{8}\sum_{ijk}(-)^{j+k}G_{ijk}^{A},  \label{parametersC} \\
a_{101} &=&\frac{1}{8}\sum_{ijk}(-)^{i+k}G_{ijk}^{A},\,\,\,\,a_{111}=\frac{1%
}{8}\sum_{ijk}(-)^{i+j+k}G_{ijk}^{A}.  \label{parametersD}
\end{eqnarray}%
Using Eq.~(\ref{eq:AlicePayoff}), we then can find the payoff for each
player 
\end{subequations}
\begin{subequations}
\label{eq:ThreePlayerPurePayoffsAlice}
\begin{align}
\Pi _{A}(\kappa _{i}^{1},\kappa _{j}^{2},\kappa _{k}^{3})& =a_{000}+\cos
\gamma \{a_{100}X_{i}+a_{010}Y_{j}+a_{001}Z_{k}\}  \notag \\
& +a_{110}X_{i}Y_{j}+a_{101}X_{i}Z_{k}+a_{011}Y_{j}Z_{k}+a_{111}\{\cos
\gamma X_{i}Y_{j}Z_{k}+\sin \gamma \Theta _{ijk}\},  \notag \\
& \\
\Pi _{B}(\kappa _{i}^{1},\kappa _{j}^{2},\kappa _{k}^{3})& =b_{000}+\cos
\gamma \{b_{100}X_{i}+b_{010}Y_{j}+b_{001}Z_{k}\}  \notag \\
& +b_{110}X_{i}Y_{j}+b_{101}X_{i}Z_{k}+b_{011}Y_{j}Z_{k}+b_{111}\{\cos
\gamma X_{i}Y_{j}Z_{k}+\sin \gamma \Theta _{ijk}\},  \notag \\
& \\
\Pi _{C}(\kappa _{i}^{1},\kappa _{j}^{2},\kappa _{k}^{3})& =c_{000}+\cos
\gamma \{c_{100}X_{i}+c_{010}Y_{j}+c_{001}Z_{k}\}  \notag \\
& +c_{110}X_{i}Y_{j}+c_{101}X_{i}Z_{k}+c_{011}Y_{j}Z_{k}+c_{111}\{\cos
\gamma X_{i}Y_{j}Z_{k}+\sin \gamma \Theta _{ijk}\},  \notag \\
&
\end{align}%
where, as Eqs.~(\ref{eq:XYZRelations}) show, the three measurement
directions $\kappa _{i}^{1},\kappa _{j}^{2},\kappa _{k}^{3}$ are held in $%
X_{i},Y_{i},Z_{i}$. Alternatively, in order to produce other quantum game
frameworks \cite{Eisert1999,MarinattoWeber}, we can interpret the rotors $A,B,C$,
held in $X_{i},Y_{i},Z_{i}$, as the unitary operations which can be applied
by each player to their qubit, where in this case, the measurement
directions will be set by the referee.

\subsubsection*{Mixed-strategy payoff relations}

For a mixed strategy game, Alice, Bob and Chris choose their first
measurement directions $\kappa _{1}^{1},$ $\kappa _{1}^{2},$ $\kappa
_{1}^{3} $ with probabilities $x$, $y$ and $z$ respectively, where $x,y,z\in
\lbrack 0,1]$ and hence choose the directions $\kappa _{2}^{1},$ $\kappa
_{2}^{2},$ $\kappa _{2}^{3}$ with probabilities $(1-x)$, $(1-y)$, $(1-z)$,
respectively. Alice's payoff is now given as 
\end{subequations}
\begin{eqnarray}
&&\Pi _{A}(x,y,z)  \notag \\
&=&xyz\sum_{i,j,k=0}^{1}P_{ijk}(\kappa _{1}^{1},\kappa _{1}^{2},\kappa
_{1}^{3})G_{ijk}+x(1-y)z\sum_{i,j,k=0}^{1}P_{ijk}(\kappa _{1}^{1},\kappa
_{2}^{2},\kappa _{1}^{3})G_{ijk}  \notag \\
&+&(1-x)yz\sum_{i,j,k=0}^{1}P_{ijk}(\kappa _{2}^{1},\kappa _{1}^{2},\kappa
_{1}^{3})G_{ijk}+(1-x)(1-y)z\sum_{i,j,k=0}^{1}P_{ijk}(\kappa _{2}^{1},\kappa
_{2}^{2},\kappa _{1}^{3})G_{ijk}  \notag \\
&+&xy(1-z)\sum_{i,j,k=0}^{1}P_{ijk}(\kappa _{1}^{1},\kappa _{1}^{2},\kappa
_{2}^{3})G_{ijk}+x(1-y)(1-z)\sum_{i,j,k=0}^{1}P_{ijk}(\kappa _{1}^{1},\kappa
_{2}^{2},\kappa _{2}^{3})G_{ijk}  \notag \\
&+&(1-x)y(1-z)\sum_{i,j,k=0}^{1}P_{ijk}(\kappa _{2}^{1},\kappa
_{1}^{2},\kappa
_{2}^{3})G_{ijk}+(1-x)(1-y)(1-z)\sum_{i,j,k=0}^{1}P_{ijk}(\kappa
_{2}^{1},\kappa _{2}^{2},\kappa _{2}^{3})G_{ijk}.  \notag \\
&&  \label{eq:AlicePayoffFourCoin}
\end{eqnarray}

\subsubsection*{Payoff relations for a symmetric game}

For a symmetric game we have $\Pi _{A}(x,y,z)=$ $\Pi _{A}(x,z,y)=$ $\Pi
_{B}(y,x,z)=$ $\Pi _{B}(z,x,y)=$ $\Pi _{C}(y,z,x)=$ $\Pi _{C}(z,y,x)$. This
requires $a_{111}=b_{111}=c_{111},$ $a_{000}=b_{000}=c_{000},$ $%
a_{110}=b_{110}=a_{101}=c_{101}=b_{011}=c_{011},$ $%
b_{100}=c_{100}=a_{010}=c_{010}=a_{001}=b_{001},$ $a_{100}=b_{010}=c_{001}$
and $a_{011}=b_{101}=c_{110}$. The payoff relations (\ref%
{eq:ThreePlayerPurePayoffsAlice}) are then reduced to 
\begin{subequations}
\begin{eqnarray}
\Pi _{A}(\kappa _{i}^{1},\kappa _{j}^{2},\kappa _{k}^{3}) &=&a_{000}+\cos
\gamma \{a_{100}X_{i}+a_{001}Y_{j}+a_{001}Z_{k}\}  \notag \\
&+&a_{110}X_{i}\{Y_{j}+Z_{k}\}+a_{011}Y_{j}Z_{k}+a_{111}\{\cos \gamma
X_{i}Y_{j}Z_{k}+\sin \gamma \Theta _{ijk}\}, \\
\Pi _{B}(\kappa _{i}^{1},\kappa _{j}^{2},\kappa _{k}^{3}) &=&a_{000}+\cos
\gamma \{a_{001}X_{i}+a_{100}Y_{j}+a_{001}Z_{k}\}  \notag \\
&+&a_{110}Y_{j}\{X_{i}+Z_{k}\}+a_{011}X_{i}Z_{k}+a_{111}\{\cos \gamma
X_{i}Y_{j}Z_{k}+\sin \gamma \Theta _{ijk}\}, \\
\Pi _{C}(\kappa _{i}^{1},\kappa _{j}^{2},\kappa _{k}^{3}) &=&a_{000}+\cos
\gamma \{a_{001}X_{i}+a_{001}Y_{j}+a_{100}Z_{k}\}  \notag \\
&+&a_{110}Z_{k}\{X_{i}+Y_{j}\}+a_{011}X_{i}Y_{j}+a_{111}\{\cos \gamma
X_{i}Y_{j}Z_{k}+\sin \gamma \Theta _{ijk}\}.
\end{eqnarray}

\subsection*{Embedding the classical game}

If we consider a strategy triplet $(x,y,z)=(0,1,0)$ for example, at zero
entanglement, then the payoff to Alice is obtained from Eq.~(\ref%
{eq:AlicePayoffFourCoin}) to be 
\end{subequations}
\begin{eqnarray}
\Pi _{A}(x,y,z) &=&\frac{1}{8}%
[G_{000}(1+X_{2})(1+Y_{1})(1+Z_{2})+G_{100}(1-X_{2})(1+Y_{1})(1+Z_{2}) 
\notag \\
&&+G_{010}(1+X_{2})(1-Y_{1})(1+Z_{2})+G_{110}(1-X_{2})(1-Y_{1})(1+Z_{2}) 
\notag \\
&+&G_{001}(1+X_{2})(1+Y_{1})(1-Z_{2})+G_{101}(1-X_{2})(1+Y_{1})(1-Z_{2}) 
\notag \\
&&+G_{011}(1+X_{2})(1-Y_{1})(1-Z_{2})+G_{111}(1-X_{2})(1-Y_{1})(1-Z_{2})].
\end{eqnarray}%
Hence, in order to achieve the classical payoff of $G_{101}$ for this
triplet, we can see that we require $X_{2}=-1$, $Y_{1}=+1$ and $Z_{2}=-1$.

This shows that we can select any required classical payoff by the
appropriate selection of $X_{i}$, $Y_{i},$ $Z_{i}=\pm 1$. Referring to Eq.~(%
\ref{eq:XYZRelations}), we therefore have the conditions for obtaining
classical mixed-strategy payoff relations as 
\begin{subequations}
\begin{eqnarray}
X_{i} &=&\cos \alpha _{1}\cos \kappa _{i}^{1}+\cos \alpha _{3} \sin \alpha
_{1}\sin \kappa _{i}^{1}=\pm 1, \\
Y_{j} &=&\cos \beta _{1}\cos \kappa _{j}^{2}+\cos \beta_{3} \sin \beta
_{1}\sin \kappa _{j}^{2}=\pm 1, \\
Z_{k} &=&\cos \chi _{1}\cos \kappa _{k}^{3}+\cos \chi _{3}\sin \chi _{1}\sin
\kappa _{k}^{3}=\pm 1.
\end{eqnarray}

For the equation for Alice, we have two classes of solution: If $\alpha
_{3}\neq 0$, then for the equations satisfying $X_{2}=Y_{2}=Z_{2}=-1$ we
have for Alice in the first equation $\alpha _{1}=0$, $\kappa _{2}^{1}=\pi $
or $\alpha _{1}=\pi $, $\kappa _{2}^{1}=0$ and for the equations satisfy $%
X_{1}=Y_{1}=Z_{1}=+1$ we have $\alpha _{1}=\kappa _{1}^{1}=0$ or $\alpha
_{1}=\kappa _{1}^{1}=\pi $, which can be combined to give either $\alpha
_{1}=0$, $\kappa _{1}^{1}=0$ and $\kappa _{2}^{1}=\pi $ or $\alpha _{1}=\pi $%
, $\kappa _{1}^{1}=\pi $ and $\kappa _{2}^{1}=0$. For the second class with $%
\alpha _{3}=0$ we have the solution $\alpha _{1}-\kappa _{2}^{1}=\pi $ and
for $X_{1}=Y_{1}=Z_{1}=+1$ we have $\alpha _{1}-\kappa _{2}^{1}=0$.

So in summary for both cases we have that the two measurement directions are 
$\pi $ out of phase with each other, and for the first case ($\alpha
_{3}\neq 0$) we can freely vary $\alpha _{2}$ and $\alpha _{3}$, and for the
second case ($\alpha _{3}=0$), we can freely vary $\alpha _{1}$ and $\alpha
_{2}$ to change the initial quantum quantum state without affecting the game
Nash equilibrium~(NE) or payoffs \cite{Rasmusen,Binmore}. The same arguments
hold for the equations for $Y$ and $Z$. Using these results in Eq.~(\ref%
{eq:thetaEquation}) we find that $\Theta _{ijk}=0$.

We have the associated payoff for Alice 
\end{subequations}
\begin{eqnarray}
\Pi _{A}(x,y,z) &=&\frac{1}{2}[G_{000}+G_{111}-\cos \gamma (G_{000}-G_{111})
\notag \\
&-&4(y+z)(a_{110}+a_{011})+\cos \gamma
\{4x(a_{111}+a_{100})+4(a_{111}+a_{001})(y+z)\}  \notag \\
&+&8xa_{110}(y+z-1)+8yza_{011}-8a_{111}\cos \gamma \{xy+xz+yz-2xyz\}].
\label{eq:AlicePayoffQuantumGameClassical}
\end{eqnarray}%
Setting $\gamma =0$ in Eq.~(\ref{eq:AlicePayoffQuantumGameClassical}) we
find Alice's payoff as 
\begin{eqnarray}
\Pi _{A}(x,y,z)
&=&G_{111}+x(G_{011}-G_{111})+y(G_{110}-G_{111})+z(G_{110}-G_{111})  \notag
\\
&&+4xy(a_{110}-a_{111})+4xz(a_{110}-a_{111})+4yz(a_{011}-a_{111})+8xyza_{111},
\notag \\
&&  \label{eq:AlicePayoffClassicalEmbedding}
\end{eqnarray}%
which has the same payoff structure as the mixed-strategy version of the
classical game.

Now, we can also write the equations governing the NE as 
\begin{eqnarray}
&&\Pi _{A}(x^{\ast },y^{\ast },z^{\ast })-\Pi _{A}(x,y^{\ast },z^{\ast }) 
\notag \\
&=&(x^{\ast }-x)[a_{110}(2y^{\ast }-1)+a_{101}(2z^{\ast }-1)+\cos \gamma
\{a_{100}+a_{111}(2y^{\ast }-1)(2z^{\ast }-1)\}]\geq 0  \notag \\
&&\Pi _{B}(x^{\ast },y^{\ast },z^{\ast })-\Pi _{B}(x^{\ast },y,z^{\ast }) 
\notag \\
&=&(y^{\ast }-y)[b_{110}(2x^{\ast }-1)+b_{011}(2z^{\ast }-1)+\cos \gamma
\{b_{010}+b_{111}(2x^{\ast }-1)(2z^{\ast }-1)\}]\geq 0  \notag \\
&&\Pi _{C}(x^{\ast },y^{\ast },z^{\ast })-\Pi _{C}(x^{\ast },y^{\ast },z) 
\notag \\
&=&(z^{\ast }-z)[c_{101}(2x^{\ast }-1)+c_{011}(2y^{\ast }-1)+\cos \gamma
\{c_{001}+c_{111}(2x^{\ast }-1)(2y^{\ast }-1)\}]\geq 0,  \notag \\
&&  \label{eq:NEEPREmbeddedStagHuntThree}
\end{eqnarray}%
where the strategy triple $(x^{\ast },y^{\ast },z^{\ast })$ is a NE. Using
the conditions defined earlier for a symmetric game, we can reduce our
equations governing the NE for the three players to 
\begin{subequations}
\label{eq:NEEPREmbeddedThreeReduced}
\begin{align}
& (x^{\ast }-x)[2a_{110}(y^{\ast }+z^{\ast }-1)+\cos \gamma
\{a_{100}+a_{111}(2y^{\ast }-1)(2z^{\ast }-1)\}]\geq 0, \\
& (y^{\ast }-y)[2a_{110}(x^{\ast }+z^{\ast }-1)+\cos \gamma
\{a_{100}+a_{111}(2x^{\ast }-1)(2z^{\ast }-1)\}]\geq 0, \\
& (z^{\ast }-z)[2a_{110}(x^{\ast }+y^{\ast }-1)+\cos \gamma
\{a_{100}+a_{111}(2x^{\ast }-1)(2y^{\ast }-1)\}]\geq 0.
\end{align}

We can see that the new quantum behavior is governed solely by the payoff
matrix through $a_{100}$, $a_{110}$ and $a_{111}$ and by the entanglement
angle $\gamma $, and not by other properties of the quantum state.

For completeness, we have Bob's payoff, in the symmetric case, as

\end{subequations}
\begin{eqnarray}
\Pi _{B}(x,y,z) &=&\frac{1}{2}[G_{000}+G_{111}-\cos \gamma (G_{000}-G_{111})
\notag \\
&-&4(x+z)(a_{110}+a_{011})+\cos \gamma \lbrack
4y(a_{111}+a_{100})+4(x+z)(a_{111}+a_{001})]  \notag \\
&+&8ya_{110}(x+z-1)+8xza_{011}-8a_{111}\cos \gamma \{xy+xz+yz-2xyz\}].
\label{eq:BobsPayoffQuantumGameClassical}
\end{eqnarray}

The mixed NE for all players is 
\begin{equation}
x^* = y^* = z^* = \frac{-a_{110}+ \cos \gamma a_{111} \pm \sqrt{a_{110}^2 -
\cos \gamma a_{100} a_{111}}}{2 \cos \gamma a_{111}} .
\end{equation}

\subsubsection*{Maximally entangled case}

For $\gamma =\pi /2$ at maximum entanglement for both NE of $(x^{\ast
},y^{\ast },z^{\ast })=(0,0,0)$ and $(x^{\ast },y^{\ast },z^{\ast })=(1,1,1)$
we have the payoff 
\begin{equation}
\Pi _{A}(x^{\ast },y^{\ast },z^{\ast })=\Pi _{B}(x^{\ast },y^{\ast },z^{\ast
})=\Pi _{C}(x^{\ast },y^{\ast },z^{\ast })=\frac{1}{2}(G_{000}+G_{111})
\end{equation}%
which gives the average of the two corners of the payoff matrix, which is as
expected.

\subsubsection*{Prisoners' Dilemma}

An example of a three-player PD game is shown in Table~\ref{ExPDTable}. For
this game, from Eq.~(\ref{eq:aDefs}), we have $a_{000}=32/8, \, a_{001}=14/8 
$, $a_{010}=14/8,\,a_{011}=0,\,a_{100}=-8/8,\,a_{101}=-2/8,\,a_{110}=-2/8,
\,a_{111}=0$, with the NE from Eqs.~(\ref{eq:NEEPREmbeddedThreeReduced})
given by 
\begin{subequations}
\begin{eqnarray}
&&(x^{\ast }-x)[-(y^{\ast }+z^{\ast }-1)-2\cos \gamma ]\geq 0, \\
&&(y^{\ast }-y)[-(x^{\ast }+z^{\ast }-1)-2\cos \gamma ]\geq 0, \\
&&(z^{\ast }-z)[-(x^{\ast }+y^{\ast }-1)-2\cos \gamma ]\geq 0.
\end{eqnarray}

% Table 1 location

We have the classical NE of $(x^{\ast },y^{\ast },z^{\ast })=(0,0,0)$ for $%
\cos \gamma =1$, but we have a phase transition, as the entanglement
increases, at $\cos \gamma =\frac{1}{2}$ where we find the new NE $(x^{\ast
},y^{\ast },z^{\ast })=(1,0,0)$, $(x^{\ast },y^{\ast },z^{\ast })=(0,1,0)$
and $(x^{\ast },y^{\ast },z^{\ast })=(0,0,1)$.
% Fig 2
The payoff for Alice from Eq.~(\ref{eq:AlicePayoffQuantumGameClassical}) is
given by 
\end{subequations}
\begin{equation}
\Pi _{A}(x,y,z)=\frac{1}{2}[7+2x+(y+z)(1-2x)-\cos \gamma \{5+4x-7(y+z)\}].
\label{eq:AlicePayoffQuantumGamePD}
\end{equation}

\begin{figure}[!ht]
\begin{center}
\includegraphics[width=3.7in]{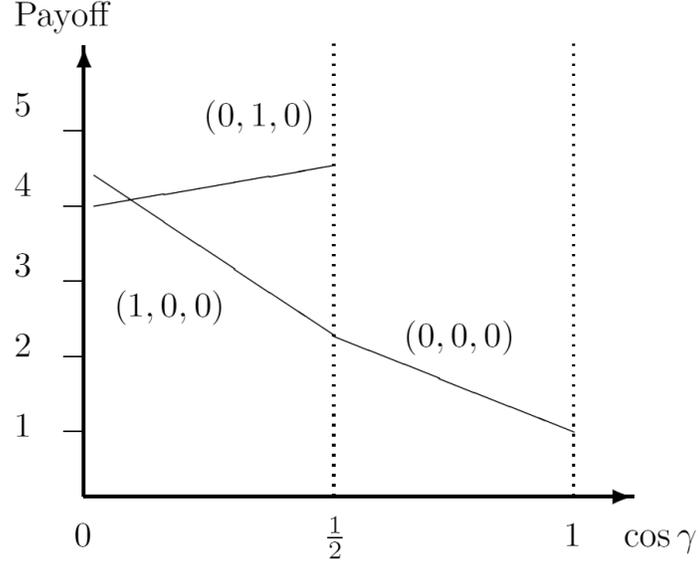}
\end{center}
\caption{{\bf{Phase structure for Alice in quantum PD game using EPR setting.}} For
the PD example given in Table~1, the classical outcome of (0,0,0), is still
returned for low entanglement, $\cos \protect\gamma >\frac{1}{2}$, but with
new NE arising at higher entanglement. As the game is symmetric, we have $%
\Pi _{A}(0,1,0)=$ $\Pi _{A}(0,0,1)$ and the NE $(0,0,1)$ is not shown.}
\label{GA3Space}
\end{figure}

For the classical region we have $\Pi _{A}(0,0,0)=\Pi _{B}(0,0,0)=\Pi
_{C}(0,0,0)=\frac{7}{2}-\frac{5}{2}\cos \gamma $, which is graphed in Fig.~%
\ref{GA3Space} along with other parts of the phase diagram. It should be
noted that $\cos \gamma $ can go negative, which will produce a mirror image
about the vertical axis of the current graph. That is for $\cos \gamma $
decreasing from $-\frac{1}{2}$ to $-1$, we have a NE of $(x^{\ast },y^{\ast
},z^{\ast })=(1,1,1)$, falling from $2.25$ down to $1$. We will also have
the NE of $(x^{\ast },y^{\ast },z^{\ast })=(1,1,0)$ and $(x^{\ast },y^{\ast
},z^{\ast })=(0,1,1)$ for $-\frac{1}{2}\cos \gamma <0$.

This graph also illustrates the value of coalitions, because if Bob and
Chris both agree to implement the same strategy, then the only NE available
for $0 < \cos \gamma < \frac{1}{2} $ for example, is $(x^*,y^*,z^*) =
(1,0,0) $. However, for a NE in the region of $\cos \gamma $ just less than
one half, both Bob and Chris receive a significantly greater payoff, of
around $4.5 $ units, as opposed to $2.5 $ for Alice, so the coalition will
receive nearly twice the payoff.

\subsection*{Players sharing the W state}

The second type of three particle entangled state \cite{Dur} is the W state 
\begin{equation}
\psi =-ABC\frac{1}{\sqrt{3}}({\iota }\sigma _{2}^{1}+{\iota }\sigma _{2}^{2}+%
{\iota }\sigma _{2}^{3}),
\end{equation}%
where once again we have used the three rotors $A$, $B$ and $C$ in order to
generalize the state as far as possible. So proceeding as for the GHZ state,
the probability that a particular state will be observed after measurement
can be found to be 
\begin{align}
P_{lmn}& =\frac{1}{24}[3+(-)^{l}X_{i}+(-)^{m}Y_{j}+(-)^{n}Z_{k}  \notag \\
&
+(-)^{l+m+n}(2(X_{i}G_{j}H_{k}+F_{i}Y_{j}H_{k}+F_{i}G_{j}Z_{k}+X_{i}V_{j}W_{k}+U_{i}Y_{j}W_{k}+U_{i}V_{j}Z_{k})-3X_{i}Y_{j}Z_{k})
\notag \\
&
+(-)^{l+m}(2F_{i}G_{j}+2U_{i}V_{j}-X_{i}Y_{j})+(-)^{l+n}(2F_{i}H_{k}+2U_{i}W_{k}-X_{i}Z_{k})
\notag \\
& +(-)^{m+n}(2G_{j}H_{k}+2V_{j}W_{k}-Y_{j}Z_{k})].
\label{eq:finalDensityWState}
\end{align}%
Clearly the same probability distribution would be found for the second type
of W state, shown in Eq.~(\ref{W2State}), because it is simply an inverse of
this state.

\subsubsection*{Obtaining the pure-strategy payoff relations}

With players sharing a W state, referring to Eq.~(\ref{eq:aDefs}), we
introduce the following notation for Alice 
\begin{equation}
a_{xyz}^{\prime }=\frac{1}{3}a_{xyz}.
\end{equation}%
Using the payoff function given by Eq.~(\ref{eq:AlicePayoff}), we then find
for Alice 
\begin{eqnarray}
&&\Pi _{A}(\kappa _{i}^{1},\kappa _{j}^{2},\kappa _{k}^{3})  \notag \\
&=&3a_{000}^{\prime }+a_{100}^{\prime }X_{i}+a_{010}^{\prime
}Y_{j}+a_{001}^{\prime }Z_{k}+a_{011}^{\prime
}(2G_{j}H_{k}+2V_{j}W_{k}-Y_{j}Z_{k})  \notag \\
&+&a_{110}^{\prime }(2F_{i}G_{j}+2U_{i}V_{j}-X_{i}Y_{j})+a_{101}^{\prime
}(2F_{i}H_{k}+2U_{i}W_{k}-X_{i}Z_{k})  \notag \\
&+&a_{111}^{\prime
}[2%
\{X_{i}G_{j}H_{k}+F_{i}Y_{j}H_{k}+F_{i}G_{j}Z_{k}+X_{i}V_{j}W_{k}+U_{i}Y_{j}W_{k}+U_{i}V_{j}Z_{k}\}-3X_{i}Y_{j}Z_{k}].
\notag \\
&&
\end{eqnarray}%
Similarly for other players, simply by switching to their payoff matrix in
place of Alices'.

Obviously for the W state there is no way to turn off the entanglement and
so it is not possible to embed a classical game, hence we now turn to a more
general state which is in a superposition of the GHZ and W type states.

\subsection*{Games with general three-qubit state}

It is noted in Ref.~\cite{Dur} that there are two inequivalent classes of
tripartite entanglement, represented by the GHZ and W states. More
specifically, Ref.~\cite{GeneralThree} finds a general three qubit pure
state 
\begin{equation}
|\psi \rangle _{3}=\lambda _{0}|000\rangle +\lambda _{1}\mathrm{e}^{i\phi
}|100\rangle +\lambda _{2}|101\rangle +\lambda _{3}|110\rangle +\lambda
_{4}|111\rangle
\end{equation}%
where $\lambda _{1},\phi \in \Re $, with $\lambda _{1}\geq 0$, $0\leq \phi
\leq \pi $ and $\sum_{j=0}^{4}\lambda _{j}^{2}=1$.

We have a $1:1$ mapping from complex spinors to GA given in Eq.~(\ref%
{eq:spinorMapping}), so we will have a general three qubit state represented
in GA as 
\begin{equation}
\psi =ABC[\lambda _{0}-\lambda _{1}\cos x{\iota }\sigma _{2}^{1}+\lambda
_{1}\sin x{\iota }\sigma _{1}^{1}+\lambda _{2}{\iota }\sigma _{2}^{1}{\iota }%
\sigma _{2}^{3}+\lambda _{3}{\iota }\sigma _{2}^{1}{\iota }\sigma
_{2}^{2}-\lambda _{4}{\iota }\sigma _{2}^{1}{\iota }\sigma _{2}^{2}{\iota }%
\sigma _{2}^{3}],
\end{equation}%
which with the rotors gives us 15 degrees of freedom.

We desire though, a symmetrical three-qubit state in order to guarantee a
fair game and so we construct 
\begin{equation}
|\psi \rangle _{3}=\rho _{0}|000\rangle +\rho _{1}(|001\rangle +|010\rangle
+|100\rangle )+\rho _{2}(|011\rangle +|101\rangle +|110\rangle )+\rho
_{3}|111\rangle
\end{equation}%
as the most general symmetrical three qubit quantum state, with $\rho _{i}$
subject to the conventional normalization conditions. We might think to add
complex phases to the four terms, however we find that this addition has no
effect on the payoff or the NE and so can be neglected. This symmetrical
state can be represented in GA, by referring to Eq.~(\ref{threeQubitBasisGA}%
), as 
\begin{eqnarray}
\psi &=&ABC[\cos \frac{\gamma }{2}\cos \frac{\phi }{2}+\sin \frac{\phi }{2}%
\sin \frac{\delta }{2}({\iota }\sigma _{2}^{1}+{\iota }\sigma _{2}^{2}+{%
\iota }\sigma _{2}^{3})/\sqrt{3}  \notag \\
&+&\sin \frac{\phi }{2}\cos \frac{\delta }{2}({\iota }\sigma _{2}^{1}{\iota }%
\sigma _{2}^{2}+{\iota }\sigma _{2}^{2}{\iota }\sigma _{2}^{3}+{\iota }%
\sigma _{2}^{1}{\iota }\sigma _{2}^{3})/\sqrt{3}+\sin \frac{\gamma }{2}\cos 
\frac{\phi }{2}{\iota }\sigma _{2}^{1}{\iota }\sigma _{2}^{2}{\iota }\sigma
_{2}^{3}].  \label{eq:GHZWCombState}
\end{eqnarray}%
If we set $\gamma =0$ and $\phi =0$ we find the product state $|000\rangle $%
, which we will constrain to return the classical game as for the GHZ state.
For $\gamma =\pi /2$ and $\phi =0$ we produce the maximally entangled GHZ
state and for $\phi =\pi $ we have the W type states in a superposition
controlled by $\delta $. Using Eq.~(\ref{eq:GHZWCombState}) and following
the same calculation path used for the GHZ state, we can arrive at the NE,
using the same condition for classical embedding as for the GHZ state,
finding for Alice 
\begin{eqnarray}
&&\Pi _{A}(x^{\ast },y^{\ast },z^{\ast })-\Pi _{A}(x,y^{\ast },z^{\ast }) 
\notag \\
&=&(x^{\ast }-x)[3(a_{100}+U_{2})\cos \gamma (1+\cos \phi )+2U_{1}(1+2\cos
\phi )-(a_{100}-3U_{2})(1-\cos \phi )\cos \delta ],  \notag \\
&&  \label{eq:GHZWCombStateNE}
\end{eqnarray}%
where 
\begin{subequations}
\begin{eqnarray}
U_{1} &=&a_{110}(2y^{\ast }-1)+a_{101}(2z^{\ast }-1)=2a_{110}(y^{\ast
}+z^{\ast }-1) \\
U_{2} &=&a_{111}(1-2y^{\ast })(1-2z^{\ast }).
\end{eqnarray}%
We can see the effect of the W type states in the $\cos \delta $ term and so
it illustrates how both types of W states contribute. The reason they can
both appear is because by demanding the classical embedding we have severely
restricted the available unitary transformations available to transform the
starting state.

\subsubsection*{The payoff relations}

The payoff function for Alice given by 
\end{subequations}
\begin{eqnarray}
\Pi _{A} &=&a_{000}-\frac{1}{2}(V_{1}+V_{3})\cos \gamma (1+\cos \phi )+\frac{%
1}{3}V_{2}(1+2\cos \phi )  \notag \\
&&+\frac{1}{6}(V_{1}-3V_{3})(1-\cos \phi )\cos \delta ,
\label{General3Payoff}
\end{eqnarray}%
where 
\begin{subequations}
\begin{align}
V_{1}& =a_{100}(1-2x)+a_{010}(1-2y)+a_{001}(1-2z) \\
V_{2}& =a_{110}(1-2x)(1-2y)+a_{101}(1-2x)(1-2z)+a_{011}(1-2y)(1-2z) \\
V_{3}& =a_{111}(1-2x)(1-2y)(1-2z).
\end{align}%
The payoff for Bob and Chris found by simply replacing $a_{ijk}$ with $%
b_{ijk}$ and $c_{ijk}$ from their respective payoff matrices. When comparing
with the payoff formula above with the classical result at $(x,y,z)=(0,0,0)$%
, it is helpful to note that $%
a_{000}+a_{001}+a_{010}+a_{011}+a_{100}+a_{101}+a_{110}+a_{111}=G_{000}$ and
generally $%
a_{000}+(-1)^{n}a_{001}+(-1)^{m}a_{010}+(-1)^{m+n}a_{011}+(-1)^{l}a_{100}+(-1)^{l+n}a_{101}+(-1)^{l+m}a_{110}+(-1)^{l+m+n}a_{111}=G_{lmn} 
$.

\subsubsection*{Uniform superposition state}

If we select a uniform superposition state, with $\rho _{0}=\rho _{1}=\rho
_{2}=\rho _{3}=\frac{1}{2}$, that is, substituting $\gamma =\frac{\pi }{2}$, 
$\phi =\frac{2\pi }{3}$ and $\delta =\frac{\pi }{2}$ , giving a product
state \textrm{H}$^{\otimes 3}|000\rangle $, with \textrm{H} being the
Hadamard operator, then we find that $\Pi _{A}(x^{\ast },y^{\ast },z^{\ast
})-\Pi _{A}(x,y^{\ast },z^{\ast })=0$ for Alice, and similarly for the other
players. That is the payoff will be independent of the player choices and
Eq.~(\ref{General3Payoff}) gives $\Pi _{A}=\Pi _{B}=\Pi _{C}=a_{000}$. Where 
$a_{000}$ represents the average of all the entries in the payoff matrix, as
expected for a uniform superposition state.

\subsubsection*{Prisoners' Dilemma}

For the PD game from the previous section with the GHZ state, we found $%
a_{100}=-8/8,a_{110}=-2/8,a_{111}=0$, so $U_{2}=0$, with the NE from Eq.~(%
\ref{eq:GHZWCombStateNE}) for the three players given by 
\end{subequations}
\begin{subequations}
\begin{align}
& (x^{\ast }-x)[(1-y^{\ast }-z^{\ast })(1+2\cos \phi )-3\cos \gamma (1+\cos
\phi )+(1-\cos \phi )\cos \delta ]\geq 0, \\
& (y^{\ast }-y)[(1-x^{\ast }-z^{\ast })(1+2\cos \phi )-3\cos \gamma (1+\cos
\phi )+(1-\cos \phi )\cos \delta ]\geq 0, \\
& (z^{\ast }-z)[(1-x^{\ast }-y^{\ast })(1+2\cos \phi )-3\cos \gamma (1+\cos
\phi )+(1-\cos \phi )\cos \delta ]\geq 0,
\end{align}%
with the payoff for Alice given by 
\end{subequations}
\begin{equation}
\Pi _{A}=4-\frac{1}{6}(1-2x)(1-y-z)(1+2\cos \phi )-\frac{1}{4}%
(5+4x-7y-7z)[\cos \gamma (1+\cos \phi )-\frac{1}{3}(1-\cos \phi )\cos \delta
].  \label{PDPayoff}
\end{equation}%
We can see with $\phi =0$ we recover the NE for the GHZ state, in Eq.~(\ref%
{eq:NEEPREmbeddedThreeReduced}).

\subsubsection*{Shifting of the NE compared to the GHZ state}

We have the classical NE of $(x^*,y^*,z^*) = (0,0,0) $ for $\cos \gamma = 1 $
and $\cos \phi = 1 $, but we can see, that once again, we have a phase
transition, as the entanglement increases, to a new NE of $(x^*,y^*,z^*) =
(1,0,0) $, $(x^*,y^*,z^*) = (0,1,0) $ and $(x^*,y^*,z^*) = (0,0,1) $.

The phase transition will be at $\cos \gamma =\frac{1}{3}(2-\cos \delta )+%
\frac{2\cos \delta -1}{3(1+\cos \phi )}$. We notice that as we increase the
weighting towards the W state, by increasing $\phi $, that it becomes easier
to make the phase transition in comparison to the pure GHZ state, that is,
we improve access to the phase transition as we introduce the weight of the $%
|011\rangle +|101\rangle +|110\rangle $ state. In fact, even at $\cos \gamma
=1$, we can achieve the NE of $(x^{\ast },y^{\ast },z^{\ast })=(1,1,1)$,
with $\phi =\pi $, giving a payoff of $3\frac{1}{3}$ units.

\subsubsection*{Maximizing the payoff}

Looking at the payoff function for Alice in Eq.~(\ref{PDPayoff}), we can
seek to maximize this function. The maximum achievable payoff is found to be 
$4.5$, which is equal to the maximum payoff found for the GHZ state, see
Fig.~\ref{GA3Space}. Thus incorporating W type states into a superposition
with the GHZ state, cannot improve the maximum payoff.

% Figure 3 location

\begin{figure}[!ht]
\begin{center}
\includegraphics[width=5.5893in]{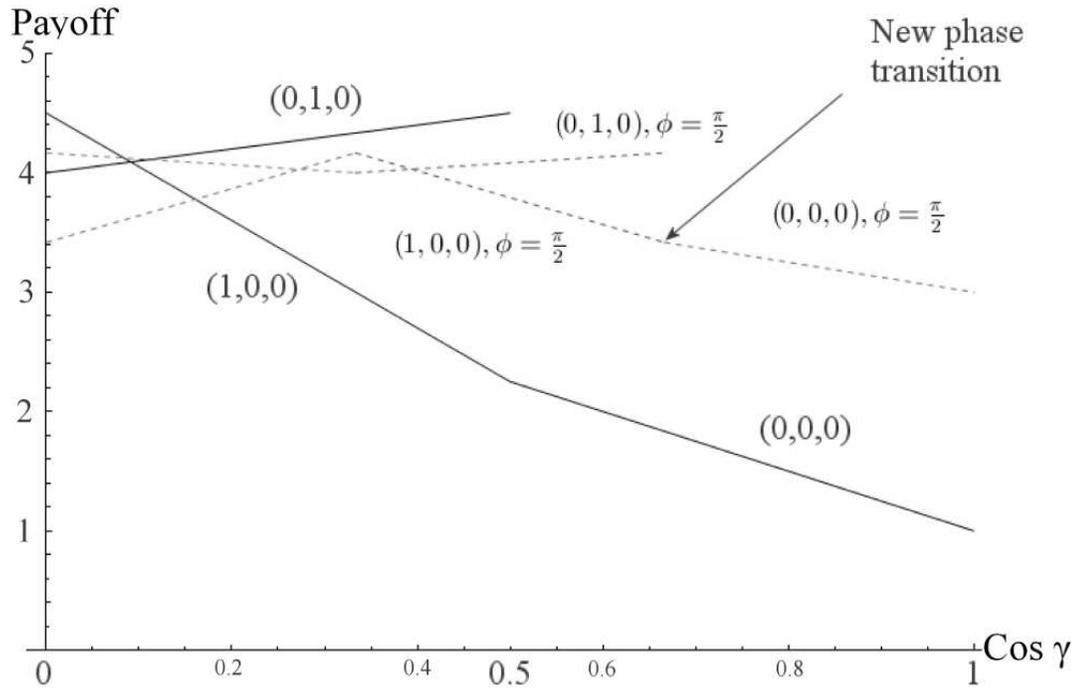}
\end{center}
\caption{{\bf{Phase transition in three-player quantum Prisoners' Dilemma with a
general three qubit state.}} The solid lines indicate the phase transitions
from Table~1, and shown in Fig.~1, with the dashed lines indicating the
shifted transitions when the W-state is mixed in. We observe that new NE now
arise at lower entanglement, at $\cos \protect\gamma = \frac{2}{3} $, as
indicated by the arrow pointer. }
\label{NewNEFigure}
\end{figure}

Observing Fig.~\ref{NewNEFigure}, we can see that as we mix in the W state,
that the phase transitions move to the right, with an extra offset available
by changing $\delta $, and the maximum payoff obtainable, will drop below
the maximum achievable of $4.5$ with the pure GHZ state. Fig.~\ref%
{NewNEFigure}, shows the shifted NE from $0.5$ to $2/3$ and payoffs for the
case $\phi =\frac{\pi }{2}$ and $\delta =0$.

\section*{Discussion}

A quantum version of a three-player two-strategy game is explored, where the
player strategy sets remain classical but their payoffs are obtained from
the outcome of quantum measurement performed, as in a typical EPR
experiment. If players share a product state, then the quantum games reduces
itself to the classical game, thus ensuring a faithful embedding of a
mixed-strategy version of a classical three-player two-strategy game within
the more general quantum version of the game.

For a general three-player two-strategy game, we find the governing equation
for a strategy triplet forming a NE is given by Eq.~(\ref{eq:GHZWCombStateNE}%
) with the associated payoff relations obtained in Eq.~(\ref{General3Payoff}%
). At zero entanglement the quantum game returns the same triplet(s) of NE
as the classical mixed-strategy game and the payoff relations in the quantum
game reduce to the trilinear form given in Eq.~(\ref%
{eq:AlicePayoffClassicalEmbedding}), equivalent to the classical game
involving mixed-strategies. We find that even though the requirement to
properly embed a classical game puts significant restrictions on the initial
quantum states, we still have a degree of freedom, available with the
entanglement angle $\gamma $, with which we can generate a new NE.

As a specific example the PD was found to have a NE of $(x^{\ast },y^{\ast
},z^{\ast })=(1,1,1)$ at high entanglement. For the GHZ state, the phase
diagram is shown in Fig.~\ref{GA3Space}, which is modulated with the
inclusion of the W type states, by reducing the payoffs and sliding the NE
closer to the classical region.

As our setup for a three-player quantum game involves players performing
classical strategies, our conclusions are restricted by not only players
sharing GHZ or W states but also by the EPR setting that we use. The most
general form of the GHZ state permits a description in terms of a single
entanglement parameter $\gamma $. However, as the general W state involves
three kets, the entanglement in such a state cannot be described by a single
parameter. It appears that as for symmetric W states with equal
superposition it is not possible to remove entanglement, therefore,
embedding a classical game within the quantum game (while players share such
states) is not possible in the EPR-type setup in which players can perform
only classical strategies. Our results in this regard are general in that
although they rely on the EPR setting, but not on a particular game as these
use the parameters introduced in Eqs. (\ref{parametersA}-\ref{parametersD})
that can be evaluated for any game. Also, this is discussed in the Section
5, where games with general three-qubit symmetric states are considered,
that include combination of GHZ and W states. However, the situation with
sharing non-equally weighted superposition states can be entirely different,
not considered in the present paper, but represents a useful extension for
future work.

Our analysis shows that, with a quantization based on the EPR setting, a
faithful embedding of a classical game can be achieved that also avoids an
Enk-Pike type argument \cite{EnkPike2002} because players' strategy sets are not extended
relative to the classical game. However, with players sharing entangled
states, while their strategy sets remain classical, our quantum games lead
to new game-theoretic outcomes.

We also find that an analysis of three-player quantum games using Clifford's
geometric algebra (GA) comes with some clear benefits, for instance, a
better perception of the quantum mechanical situation involved and
particularly an improved geometrical visualization of quantum mechanical
operations. The same results using the familiar algebra with Pauli matrices
may possibly be tractable but would certainly obscure intuition. Also, the
simple expression given in (\ref{eq:DoranOverlapProb}) for the overlap
probability between two quantum states in the $N$-particle case is another
benefit of the GA approach.

The results reported in the paper can be useful in a game-theoretic analysis
of the EPR paradox. Bell's consideration of the EPR paradox usually implies
the inconsistency between locality and completeness of quantum mechanics, or
in more broader terms, simply the surprising nonlocal effects invoked by
entanglement. However, one notices that these conclusions are merely
sufficient but not necessary for the violation of Bell's inequality and that
other interpretations are also reported \cite%
{Pitowsky,Pitowsky1,Rastall,Baere,Christian}, especially, the interpretation
based on the non-existence of a single probability space for incompatible
experimental contexts \cite{Khrennikov}. This non-existence also presents a
new route in constructing quantum games and the first step in this direction
was taken in Ref \cite{Iqbal}. Because such quantum games originate directly
from the violation of Bell's inequality, they allow a discussion of the EPR
paradox in the context of game theory. This is also supported by the fact
that for quantum games with players sharing entanglement, a game-theoretic
analysis that involves Bell's settings \cite{Bell,Bell1,Bell2} has been
reported in Refs \cite{CheonAIP,Cleve}.

A variety of other classical games could now be adapted and applied to this three-player framework, with new NE being expected.
The present study of three-player quantum games can also be naturally extended to analyze the N-player
quantum games. We believe that the mathematical formalism of GA permits this
in a way not possible using the usual complex matrices. Also, this
extension could be fruitfully exploited in developing a game-theoretic
perspective on quantum search algorithms and quantum walks.
We find that our analysis can be helpful in providing an
alternative viewpoint (with emphasis on underlying geometry) on multi-party
entanglement shared by a group of individuals (players), while they have
conflicting interests and can perform only classical actions on the quantum
state. That is, a viewpoint that is motivated by the geometrical perspective
that Clifford's geometric algebra provides. Such situations take place in
the area of quantum communication and particularly in quantum cryptography 
\cite{LeeJohnson,Pirandola:2005,houshmand2010game}.

% You may title this section "Methods" or "Models". 
% "Models" is not a valid title for PLoS ONE authors. However, PLoS ONE
% authors may use "Analysis" 

% Do NOT remove this, even if you are not including acknowledgments

\section*{Acknowledgments}

%\section*{References}
% The bibtex filename
\bibliographystyle{plain}
\bibliography{precision}

%\section*{Figure Legends}

%\begin{figure}[!ht]
%\begin{center}
%%\includegraphics[width=4in]{figure_name.2.eps}
%\end{center}
%\caption{
%{\bf Bold the first sentence.}  Rest of figure 2  caption.  Caption 
%should be left justified, as specified by the options to the caption 
%package.
%}
%\label{Figure_label}
%\end{figure}

\section*{Tables}

\begin{table}[!ht]
\caption{{\bf{An example of three-player Prisoners' Dilemma.}} }
\begin{tabular}{|l|l|l|l|l|l|l|l|l|}
\hline
State & $\ $\ $|000\rangle $ & $\ $\ $|001\rangle $ & $\ $\ $|010\rangle $ & 
$\ $\ $|100\rangle $ & $\ $\ $|011\rangle $ & $\ $\ $|101\rangle $ & $\ $\ $%
|110\rangle $ & $\ $\ $|111\rangle $ \\ \hline
Payoff & $(6,6,6)$ & $(3,3,9)$ & $(3,9,3)$ & $(9,3,3)$ & $(0,5,5)$ & $%
(5,0,5) $ & $(5,5,0)$ & $(1,1,1)$ \\ \hline
\end{tabular}%
\begin{flushleft}The payoff for each player (one,two,three), for each measurement outcome.
\end{flushleft}
\label{ExPDTable}
\end{table}

%\begin{table}[!ht]
%\caption{\bf{Table title}}
%\begin{tabular}{|c|c|c|}
%table information
%\end{tabular}
%\begin{flushleft}Table caption
%\end{flushleft}
%\label{tab:label}
% \end{table}

\end{document}